\documentclass[%
 aip,
 amsmath,amssymb,
 reprint,%
]{revtex4-1}

\usepackage{graphicx}% Include figure files
\usepackage{dcolumn}% Align table columns on decimal point
\usepackage{bm}% bold math
\usepackage[utf8]{inputenc}
\usepackage[T1]{fontenc}
\usepackage{mathptmx}
\usepackage{etoolbox}
\usepackage{xcolor}
\usepackage{ulem}

\makeatletter
\def\@email#1#2{%
 \endgroup
 \patchcmd{\titleblock@produce}
  {\frontmatter@RRAPformat}
  {\frontmatter@RRAPformat{\produce@RRAP{*#1\href{mailto:#2}{#2}}}\frontmatter@RRAPformat}
  {}{}
}%
\makeatother
\begin{document}

%\preprint{AIP/123-QED}

\title{Metastability of multi-population Kuramoto-Sakaguchi oscillators}
% Force line breaks with \\
\author{Bojun Li}
\author{Nariya Uchida}%
 \email{uchida@cmpt.phys.tohoku.ac.jp}
\affiliation{ 
Department of Physics, Tohoku University, 
Sendai 980-8578 , Japan
}%

\date{\today}% It is always \today, today,

\begin{abstract}
An Ott-Antonsen reduced $M$-population of Kuramoto-Sakaguchi oscillators is investigated, focusing on the influence of the phase-lag parameter $\alpha$ on the collective dynamics.
For oscillator populations coupled on a ring, we obtained a wide variety of spatiotemporal patterns, including coherent states, traveling waves, partially synchronized states, modulated states, and incoherent states.  
Back-and-forth transitions between these states are found, which suggest metastability.
Linear stability analysis reveals the stable regions of coherent states with different winding numbers $q$. Within certain $\alpha$ ranges, the system settles into stable traveling wave solutions despite the coherent states also being linearly stable.
For around $\alpha \approx 0.46\pi$, the system displays the most frequent metastable transitions between coherent states and partially synchronized states, while for $\alpha$ closer to $\pi/2$, metastable transitions arise between partially synchronized states and modulated states. 
This model captures metastable dynamics akin to brain activity, offering insights into the synchronization of brain networks.
\end{abstract}

\maketitle

\begin{quotation}
The rhythmic synchronization across various dynamical units is a ubiquitous phenomenon, observable in systems as diverse as swinging pendulum clocks and firing neurons in the brain. 
A characteristic behavior of brain dynamics is the chimera state where the synchronous and asynchronous regions coexist.
The chimera state is originated from the frustrated coupling which is typically attributed to the phase lag.

Our study is motivated by the hierarchical structure of the brain where billions of neurons are divided into many populations. We explore a model composed of many populations, focusing particularly on an intriguing question - how the collective behavior of a ring of coupled oscillator populations evolves as the phase lag parameter varies.
Strikingly, the system displays various distinct spatiotemporal patterns, ranging from coherence to traveling waves, partially synchronized states, modulated states, and even brain-like metastable dynamics, where the oscillators alternate between different states. These findings pave the way for new insights into synchronization on hierarchical networks, shedding light on the complex rhythmic phenomena observed in the brain.
\end{quotation}

\section{Introduction}
The collective behavior emerging from large ensembles of coupled oscillators serves as a paradigmatic model for understanding synchronization phenomena across diverse natural and engineered systems~\cite{pikovsky2001synchronization}.
{
A prominent example is the Kuramoto model~\cite{Kuramoto1975},
which was later generalized to the Kuramoto-Sakaguchi (KS) model \cite{sakaguchi1986a}
that introduces an additional phase-lag parameter.
A non-local version of the KS model led to 
Kuramoto and Battogtokh's seminal discovery~\cite{kuramoto2002} of 
chimera states chracterized by 
spatial coexistence of synchronized and desynchronized states.
Since then,
}
extensive studies have explored 
chimera states under various coupling schemes 
(for reviews, see Refs.~\cite{kuramoto2006mean,Panaggio2015, Bera2017Chimera, Omelchenko2018The, Parastesh2021Chimera}).
In addition to the original single-population case, researchers have explored the systems with both two populations \cite{Okuda1991Mutual, Montbrio2004Sync, Barreto2008Synchro, Abrams2008Solvable,Sheeba2008Routes,Sheeba2009Asymmetry,kawamura2010identical,kawamura2010nonidentical,Martens2016Chimera, Martens2016Basins}, and more than two populations \cite{Martens2010Chimeras, Martens2010Bistable,Wildie2012Metastability, Shanahan2010Metastable}. 

{
Analysis of chimera states in coupled oscillator populations often employs
a powerful dimensionality reduction technique 
called the Ott-Antonsen (OA) ansatz~\cite{ott2008low}.
It enables capturing the system's behavior using just a few mean-field variables, 
making the analysis of large populations tractable.
It has been applied to, for example, studying 
the stability of twisted states~\cite{omel2014partially}, 
multicluster chimera states~\cite{Xie2014Multi},
and emergence of turbulence~\cite{wolfrum2016turbulence}
in nonlocally coupled oscillators. 
}

Beyond its intrinsic interest, the oscillator model also holds great potential for applications in neuroscience. Neural systems have been shown to exhibit synchrony and chimera patterns at various levels \cite{MAJHI2019100}.
{Although the origin of the chimera states in brain dynamics is not elucidated yet, 
recent work~\cite{Clusella2022kuramoto} 
derived the KS model from a population of weakly coupled, nearly identical quadratic 
integrate-and-fire (QIF) neurons, and found a relationship between 
the phase lag $\alpha$ and the electrical and chemical coupling strengths.
This suggests that we may possibly regard the non-local KS model, 
which generates chimera states by tuning the phase lag close to $\pi/2$,
as a minimal model of brain dynamics instead of 
more realistic models of spiking neurons. 
}
At the macroscopic scale, the cerebral cortex is divided into many interconnected brain regions, each containing a large number of interacting neurons. These regions are connected through white matter tracts, forming a network of networks \cite{Zhou2006Hierarchical}. 
If individual neurons are represented as Kuramoto-Sakaguchi oscillators, 
then brain regions can naturally be considered populations of oscillators, 
where the OA reduction is highly applicable.
This sparked our interest in the OA-reduced $M$-population {KS} model. 
While a few previous studies have explored this model~\cite{Skardal2012Hierarchical, Smirnov2018Solitary, Laing2023chimeras}, the effects of the phase lag parameter $\alpha$ remain underexplored. 

In this paper, we consider the OA-reduced $M$-population KS model and investigate the influence of the phase lag parameter $\alpha$ on the system's collective dynamics. We examine a ring lattice, where each site hosts a population of oscillators, and find that this system exhibits metastable behavior, a hallmark of brain dynamics \cite{Emmanuelle2014the}.
Our key findings reveal that varying the phase lag parameter $\alpha$ induces intricate spatiotemporal patterns and metastability in the system's collective dynamics. These results underscore the importance of the phase lag parameter in shaping the emergent behavior of coupled oscillator networks, offering insights into the potential role of this parameter in the complex dynamics observed in the brain.

\begin{figure*}[htbp]
\includegraphics[width=17.2cm]{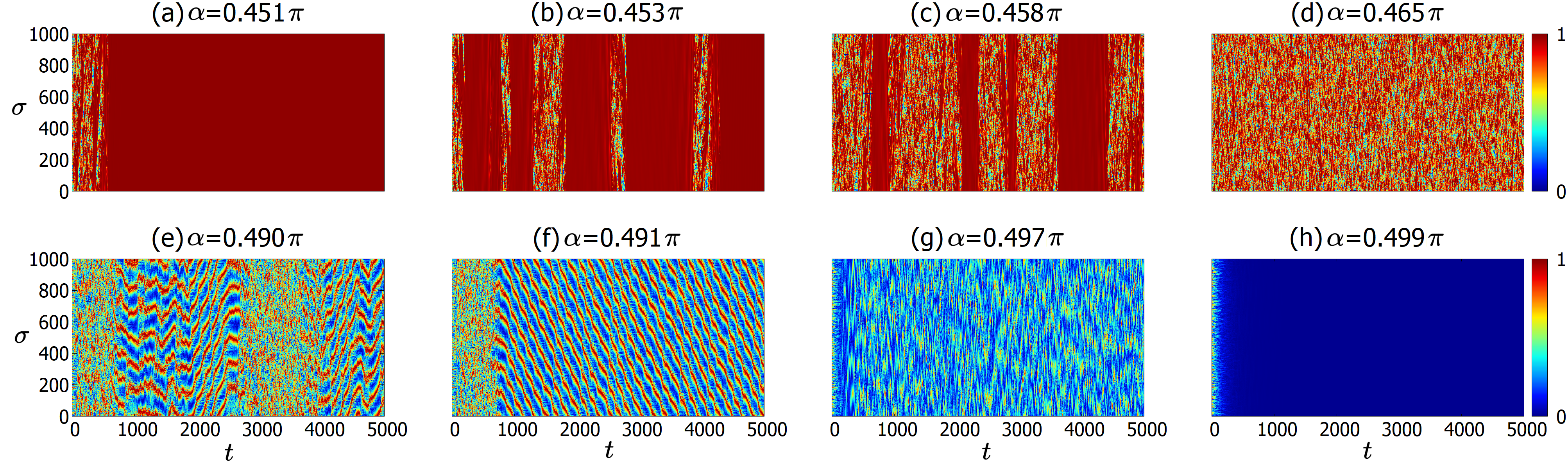}
\caption{Spatiotemporal patterns of amplitudes $r_\sigma(t)$. The vertical axis represents the index of the oscillators and the horizontal axis shows the time evolution. (a) Coherent state. (b)(c) Transitions between coherent state and partially synchronized states. (d) Partially synchronized state. (e) Transitions between partially synchronized state and modulated state. (f) Modulated state. (g)(h) The degree of incoherence increases and finally reaches a complete incoherent state.}
\label{fig:1}
\end{figure*}

\section{Model}
We consider a $M$-population of KS model, each containing $N$ oscillators,
\begin{equation}
    \frac{\mathrm{d} \theta_k^\sigma}{\mathrm{d} t} = \omega_k^\sigma - \sum_{\tau=1}^M \frac{K_{\sigma\tau}}{N} \sum_{l=1}^N \sin(\theta_k^\sigma - \theta_l^\tau + \alpha_{\sigma\tau}), \label{eq:m-population_KS}
\end{equation}
where $\sigma$ is the index of population and $k$ is the index of oscillators in a population. If we consider the thermodynamic limit $N \to \infty$, the ensemble dynamics can be represented by the oscillator density $f^\sigma(\theta,\omega)$, and then be reduced to a low-dimension mean-field dynamics via the OA ansatz \cite{ott2008low} (see Appendix~\ref{appendix_OA} for details). The complex order parameter is
\begin{equation}
    \frac{\mathrm{d} z_\sigma^*}{\mathrm{d} t}=-(\Delta_\sigma + i\Omega_\sigma) z_\sigma^* + \sum_{\tau=1}^M \frac{K_{\sigma\tau}}{2} \left [ e^{i\alpha_{\sigma\tau}} z_\tau^* - e^{-i\alpha_{\sigma\tau}} z_\tau (z_\sigma^*)^2 \right ],
    \label{eq:m-population_KS_op_dynamics}
\end{equation}  
where $\Omega_\sigma$ and $\Delta_\sigma$ are the center and width of Lorentzian distribution of natural frequency in a population.
This can be further expressed in polar coordinates defined via $z_\sigma=r_\sigma e^{-i\phi_\sigma}$ as
\begin{equation}
    \frac{\mathrm{d} r_\sigma}{\mathrm{d} t} =
    -\Delta_\sigma r_\sigma 
    + \frac{1-r_\sigma^2}{2} \sum_{\tau=1}^M K_{\sigma\tau} r_\tau \cos(\alpha_{\sigma\tau}+\phi_\tau-\phi_\sigma), \label{eq:r}
\end{equation}
\begin{equation}
    \frac{\mathrm{d} \phi_\sigma}{\mathrm{d} t} = -\Omega_\sigma + \frac{1+r_\sigma^2}{2r_\sigma} \sum_{\tau=1}^M K_{\sigma\tau} r_\tau \sin(\alpha_{\sigma\tau}+\phi_\tau-\phi_\sigma), \label{eq:phi}
\end{equation}
where $r_\sigma$ and $\phi_\sigma$ are the radial and angular components of order parameters respectively. Two population version of this equation are studied in the previous work \cite{Abrams2008Solvable, Martens2016Chimera}.

We consider the situation in which each oscillator population is placed on a ring. If there are $M$ populations,
Eq.~(\ref{eq:r})~(\ref{eq:phi}) has $M+1$ trivial solutions: 

(a) $M$ coherent solutions where all $r_\sigma$ are identical,
\begin{eqnarray}
r_\sigma &=& r,  \label{eq:5}\\
\phi_\sigma (t) &=& \phi_\sigma(0) + \omega t, \label{eq:6}\\
\phi_\sigma(0) &=& \phi_1(0) + \frac{2\pi q}{M} (\sigma-1), \label{eq:7}
\end{eqnarray}
where $q$ is the integer twist number. For simplicity, we set $M$ even, and then $q$ can take $M$ different values $q=-\frac{M}{2},-\frac{M}{2}+1,\dots,-1,0,1,\dots,\frac{M}{2}-1$. $q=0$ corresponds to the phase synchronized states and $q \ne 0$ corresponds to the twisted states.
We use a top-hat function as the coupling scheme, i.e. $K_{\sigma\tau}=K$ inside the coupling range $R$, otherwise $K_{\sigma\tau}=0$. We set identical phase lag and natural frequency distribution $\alpha_{\sigma\tau} = \alpha, \Delta_\sigma = \Delta$, and $\Omega_\sigma = 0$.
Under this setting, the solution can be calculated as
\begin{eqnarray}
r  &=& \sqrt{1-\frac{2\Delta}{Kh(q;R,M)\cos \alpha}},  \label{eq:r_solution} \\
\omega 
&=& Kh(q;R,M)\sin\alpha-\Delta\tan\alpha,  \label{eq:phi_solution}
\end{eqnarray}
where 
\begin{eqnarray}
h(q;R,M) =\begin{cases}
 2R+1 & \text{ if } q=0, \\
 \frac{\sin \frac{\pi q(2R+1)}{M}  }{\sin\frac{\pi q}{M}}  & \text{ if } q\ne0.
\end{cases}  
\end{eqnarray}
(b) 1 incoherent solution where $r_\sigma=0$. Besides, there also exist many nontrivial solutions that cannot be written in an explicit form.

From prior research, we know that in the traditional Kuramoto-Sakaguchi phase oscillator model, when the phase lag $\alpha$ is sufficiently smaller than $\pi/2$, the system will be attracted to a coherent steady state. However, as $\alpha$ approaches $\pi/2$, a chimera state emerges where the oscillator phases separate into coherent and incoherent groups. In the OA reduced $M$-population KS model, when $\alpha$ is much smaller than $\pi/2$, a similar situation arises where the system enters a coherent steady state (either trivial solutions Eq.~(\ref{eq:5})(\ref{eq:6})(\ref{eq:7}) or nontrivial solutions described in the next section). However, as $\alpha$ approaches $\pi/2$, the system exhibits significantly different behavior due to the amplitude $r$.

\section{Spatiotemporal Pattern}
The following numerical results are obtained for the parameters $M=1000, R=40, \Delta=0.01, K=0.045$. The fourth order Runge-Kutta method with a time step of 0.01 is used for time integration.
Fig.~\ref{fig:1} illustrates the spatiotemporal patterns of $r_\sigma(t)$ for different $\alpha$. Fig.~\ref{fig:1}(a) shows that for small $\alpha$, from the random initial condition, the system directly settles into a coherent state (complete synchronization or twisted states). In panels (b) and (c), we observe that this system exhibits metastable behavior: In the red region, the amplitudes $r_\sigma$ of the oscillators are nearly equal and very close to the coherent solution Eq.~(\ref{eq:r_solution}), while in the other time intervals, the oscillators are partially synchronized. Here, a partially synchronized state refers to the state where different populations exhibit varying degrees of synchrony ($r_\sigma$), and both $r_\sigma$ and $\phi_\sigma$ show a disordered spatial arrangement. 
As the parameter $\alpha$ increases, the system spends an increasingly larger fraction of time in the partially synchronized state, until only the partially synchronized state remains (Fig.~\ref{fig:1}(d)). With a further increase in $\alpha$, the system shows another kind of metastable behavior, transitioning between the partially synchronized state and the modulated state (Fig.~\ref{fig:1}(e)). In the modulated state, the system is characterized by spatially structured synchrony, where contiguous groups of populations having $r_\sigma \approx r$ are interspersed with populations having $r \approx 0$. This arrangement results in a distinct pattern of alternation between synchronized and unsynchronized populations which we call a modulated state. The characteristic width of the spatial periodic arrangement increases with the coupling range $R$.  
Finally, when $\alpha$ increases very close to $0.5\pi$, this metastable behavior will disappear again and the system becomes completely incoherent where all $r_\sigma = 0$ (Fig.~\ref{fig:1}(h)).

\section{traveling wave solution}
\begin{figure}[htbp]
\includegraphics[width=8.6cm]{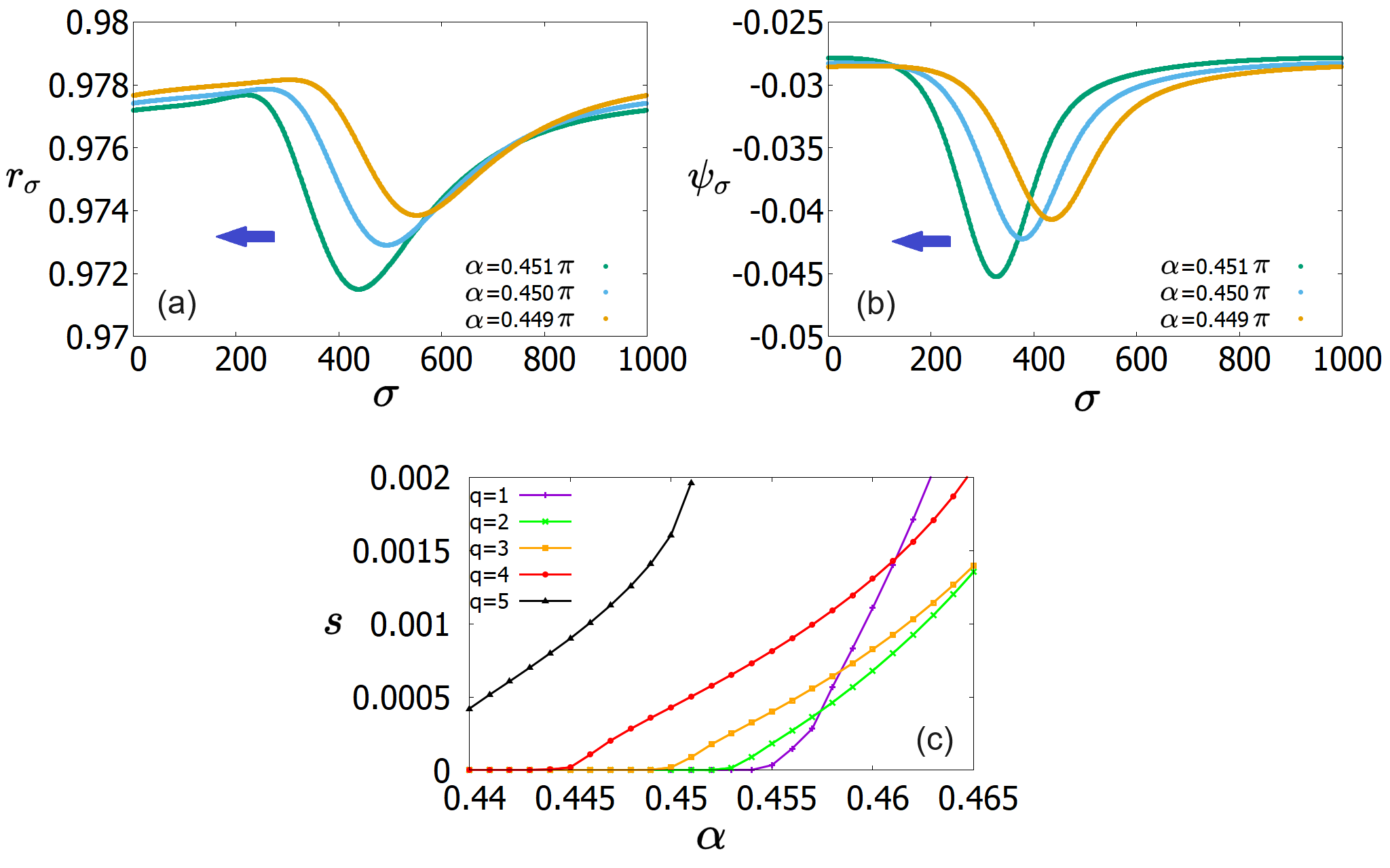}
\caption{Traveling wave solutions with $q=-5$ for $\alpha=0.449\pi,0.450\pi,0.451\pi$.
{Spatial profiles of (a) the amplitude $r_\sigma$ and (b) phase difference $\psi_\sigma$ 
between nearest neighbors.} 
The arrow shows the traveling direction of the wave. (c) Standard deviation $s$ of the amplitude {as a function of $\alpha$}. 
For $s=0$, the system is in a trivial coherent state. For $s>0$, the system exhibits a traveling wave. The threshold varies for different $q$.
}
\label{fig:2}
\end{figure}

For small values of $\alpha$, the system reaches the trivial coherent state described by Eq.~(\ref{eq:5})-(\ref{eq:7}). Depending on the initial conditions, the system reaches a stable coherent state with a particular value of the parameter $q$. However, as $\alpha$ is increased, the system will reach non-trivial steady states.
These non-trivial solutions have values close to the coherent solution but show nonuniform spatial distributions, and they propagate at constant velocities, as shown in Fig.~\ref{fig:2}(a)(b). Instead of the phase variable $\phi_\sigma$, we use the phase difference 
${\psi_\sigma}=\phi_{\sigma+1}-\phi_\sigma$ to more intuitively represent the system state. 
If the sign of the parameter $q$ is changed, the propagation direction of the waves will be reversed.
For different values of $q$, the system exhibits different thresholds of $\alpha$ above which the traveling wave state emerges. This is shown in Fig.~\ref{fig:2}(c), using the standard deviation of $r_\sigma$. Once this threshold is exceeded, further increasing $\alpha$ leads to an increase in the amplitude of the waves.
\begin{figure}[htbp]
\includegraphics[width=6.45cm]{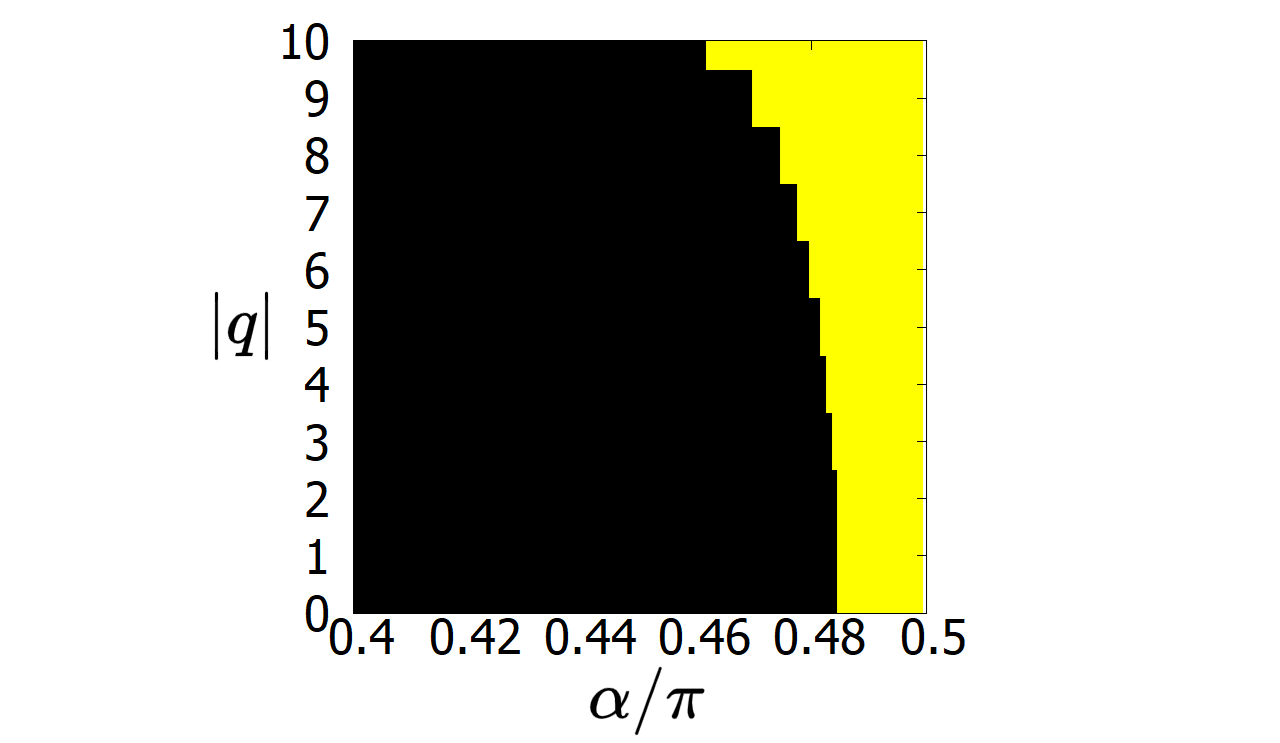}
\caption{Linear stability diagram of the coherent state with $|q|=0,1,\dots,10$. The black area shows the parameter range where the coherent state is neutrally stable. The yellow area shows where the coherent state is unstable.}
\label{fig:3}
\end{figure}

We performed a linear stability analysis of the coherent states (see Appendix~\ref{appendix_LSA}). With the fixed parameters mentioned above and varying $\alpha$, the linearly stable ranges of the different coherent states are shown in Fig.~\ref{fig:3}. In the black area, $2M-1$ eigenvalues have negative real parts, except only 1 eigenvalue having a zero real part. In the yellow area, at least one eigenvalue has a positive real part. 
It is noteworthy that within the parameter range where the system displays traveling wave solutions, the coherent solutions are also linearly stable.
However, extensive simulations show that even when both solutions are stable, all initial conditions eventually converge to the traveling wave solutions. This suggests the traveling wave solutions have a larger basin of attraction compared to the coherent solutions with the same $q$ value.

\section{Metastability}
\begin{figure}[htbp]
\includegraphics[width=8.6cm]{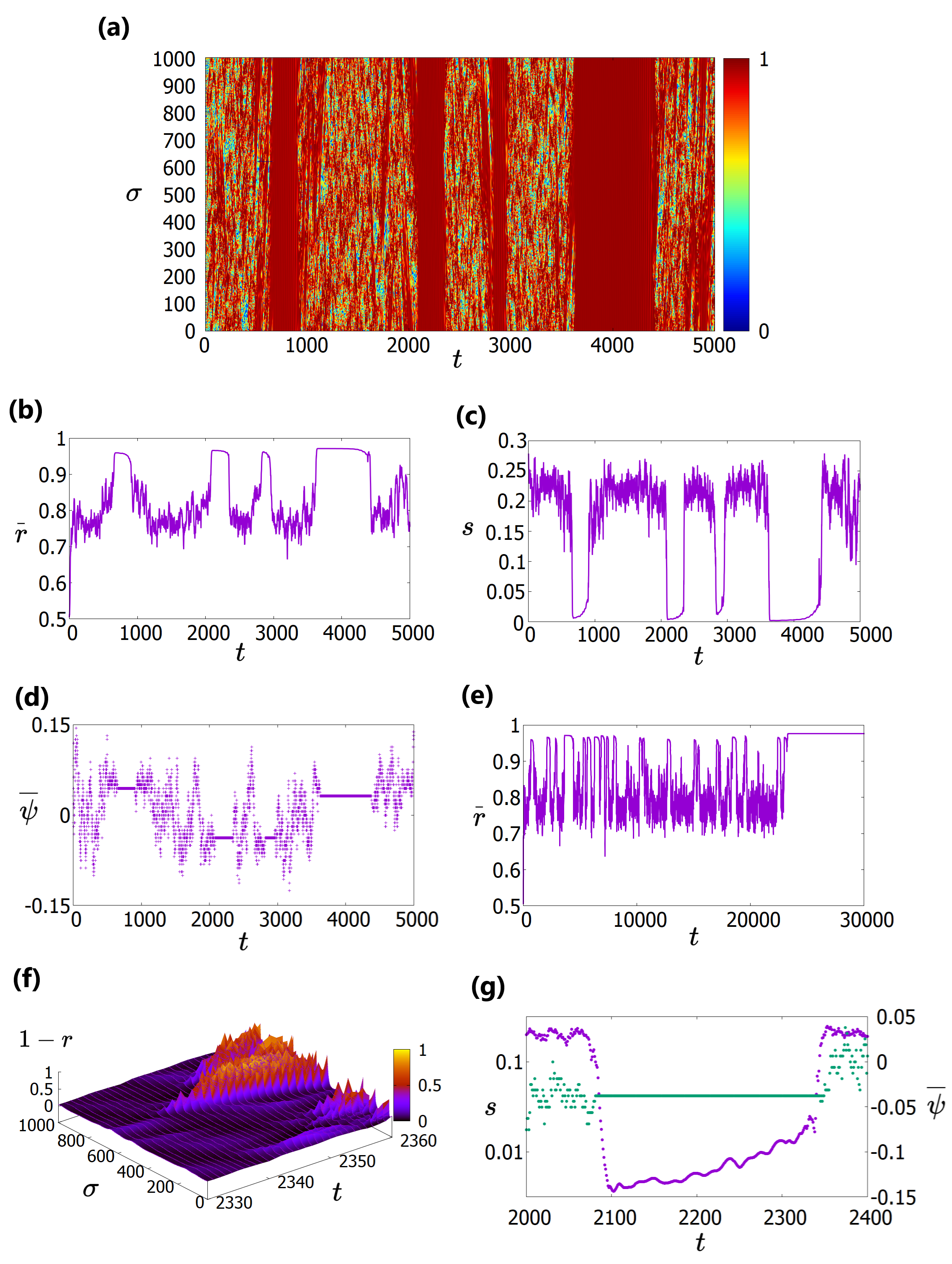}
\caption{Transition between coherent and partially synchronized states for $\alpha=0.458\pi$. (a) Spatiotemporal pattern of $r_\sigma$. 
{
In the time window $t\in[0,5000]$, the system has experienced 
4 coherent states: [656,904], [2081,2344], [2817,2956], [3622,4381]. 
Time evolution of the 
(b) spatially averaged amplitude $\overline r(t)$ and 
(c) standard deviation of amplitude $s$ for $t<5000$. 
(d) 
Time evolution of the average phase difference of neighbor oscillators 
$\overline{\psi}(t)$. 
It is constant in the coherent state and the corresponding {twist number} 
is $q=\overline{\psi}M/2\pi$. In this example, the 4 coherent states are $q=7,-6,-6,5$ respectively.
(e) 
$\overline r(t)$ on a longer timescale. 
The system eventually enters a stable coherent state at $t\approx 23260$. 
(f)
Spatiotemporal pattern of $1- r_\sigma$ in the time window [2330:2360]
showing the collapse of the coherent state. 
(g)
Time evolution of 
the standard deviation of the amplitude $s$ (logscale, purple)
and
the average phase difference $\overline{\psi}$ (green) 
in the time window [2000:2400]
showing the transitions between the incoherent and coherent states.
}
}
\label{fig:4}
\end{figure}

As mentioned earlier, before reaching the steady state, the system can undergo transitions between different metastable states. Fig.\ref{fig:4} illustrates an example for $\alpha=0.458\pi$. 
In the first $5000$ time units, the system enters and exits the coherent state 4 times
{as seen in the spatiotemporal plot of the amplitude in Fig.\ref{fig:4}(a).
We show the time evolution of the spatial average of the amplitude $r$ 
in Fig.\ref{fig:4}(b), and its standard deviation $s$ in Fig.\ref{fig:4}(c).
}
In {each coherent period},
the spatially averaged amplitude of the oscillators is close to the value given by Eq.~(\ref{eq:5}), and the average phase difference {$\overline{\psi}$} 
remains constant 
 (Fig.\ref{fig:4}{(d)}). 
{Because the latter takes the discrete values $\overline{\psi} = 2\pi q/M$,
we can unambiguously identify each coherent period by its constantness, and} count the number of {such} contiguous segments 
to determine how many transitions the system has undergone. In the example shown in Fig.\ref{fig:4}, the system experiences dozens of such transitions before ultimately settling into a stable state{, as shown in the plot of $\overline{r}$ 
on a longer timescale [Fig. \ref{fig:4}(e)].
It is important to note that 
the order parameter $r$ does not fully homogenize 
when the system enters a coherent state,
and traveling waves with a small amplitude remain.
In contrast to the stable traveling waves found 
at smaller values of $\alpha$ (Fig.\ref{fig:2}), 
the traveling waves are gradually amplified
as shown in Fig.\ref{fig:4}(f).
Once they grow sufficiently, the coherent state becomes destabilized, 
leading to the transition to the incoherent state.
In Fig.\ref{fig:4}(g), we show time evolution of the standard deviation $s$ 
of the amplitude and the average phase difference $\overline{\psi}$
during a transition from the incoherent to coherent  states and 
a successive backward transition.
It indicates that the coherent state emerges quickly, and
the standard deviation $s$ drops to below $10^{-2}$ 
within several tens of time units,
while the amplification of the traveling waves take much 
longer time where $s$ grows roughly exponentially. 
After the exponential stage, 
a sharp increase of $s$
and dehomogenization of the phase difference $\psi$ take place simultaneously.
By comparing similar plots for different coherent periods,
we find that the coherence lasts longer if 
the initial value of $s$ (after a transition from the incoherent state) is smaller, 
and that the coherence is lost when $s$ rises above a threshold in [0.05:0.1]  
(data not shown). 
The initial value of $s$ is random and thus the duration of each coherent state also has large variations.
If the initial value of $s$ is sufficiently small, 
the remaining traveling waves do not grow and the system enters a stable coherent state. 
}

\begin{figure}[htbp]
\includegraphics[width=8.6cm]{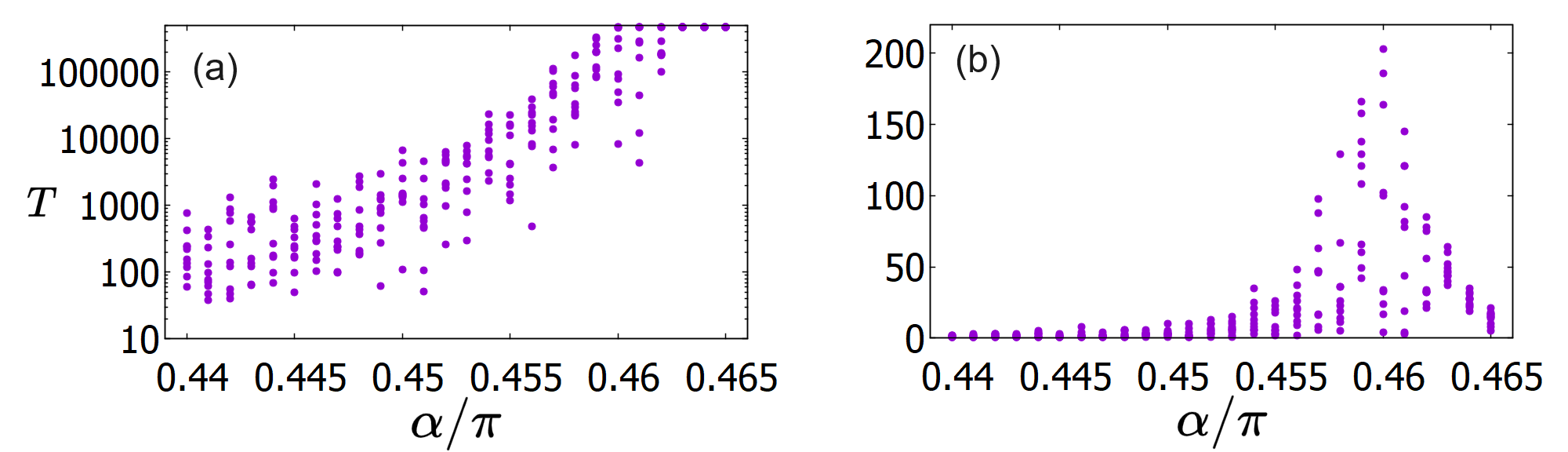}
\caption{For each $\alpha$, 10 different initial conditions were taken, and the simulations were run up to $t=500000$. (a) The lifetime of the transition process. (b) The number of coherent states system goes through within 500000 time units. 
We {judge}  that {the system has entered a coherent state} 
if $\overline{\psi}(t)$ remains constant for more than 25 time units.
This choice of threshold does not qualitatively change the result.
}
\label{fig:5}
\end{figure}

The lifetime for the system to reach a stable coherent state
for different values of $\alpha$ and the number of transitions between different coherent 
states during this process are shown in Fig.~\ref{fig:5}. 
{As shown in Fig.~\ref{fig:5}(a),
the lifetime of the metastable state increases roughly exponentially as $\alpha$ increases,
although it has a large fluctuation and depends on the initial condition.}
It can be observed that when $\alpha$ exceeds $0.46\pi$, there are no realizations with a lifetime less than 500000 time units. 
Also, for $\alpha>0.46\pi$, the number of transitions starts to decrease. 
This is because fluctuations begin to dominate the system, causing the attraction towards the coherent state to weaken.
Around $\alpha = 0.46\pi$, the system exhibits the strongest metastability. 

\begin{figure}[htbp]
\includegraphics[width=6.45cm]{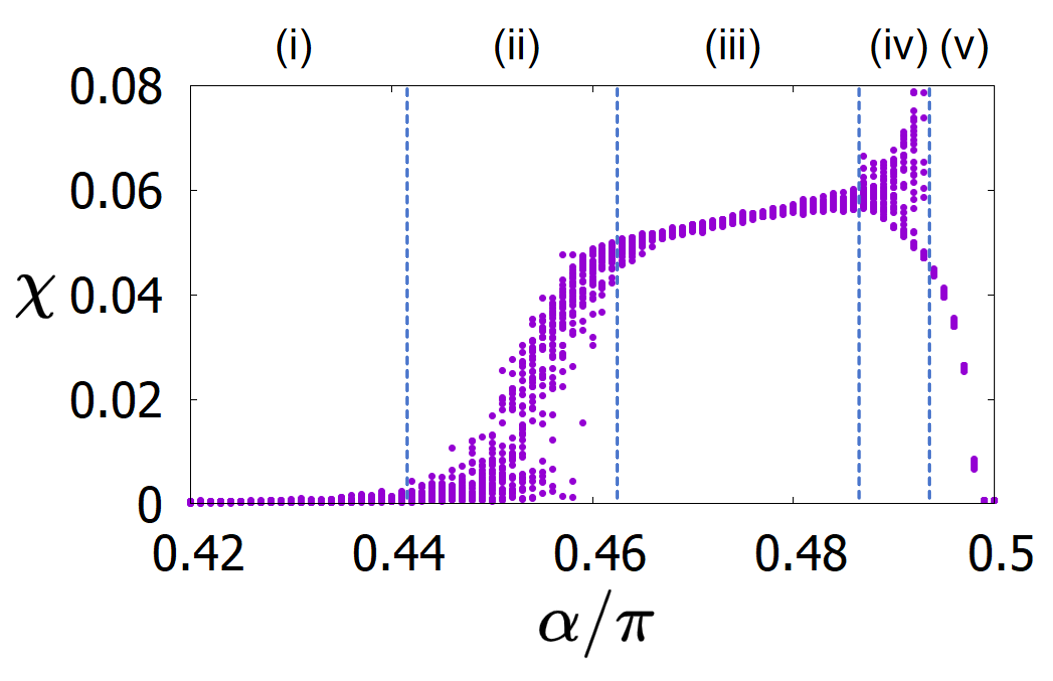}
\caption{Time average of the spatial variance of the amplitude $\chi$. Different states of the system are characterized by $\chi$. The dotted lines show $\alpha_1$, $\alpha_2$, $\alpha_3$ and $\alpha_4$. (i) $0 \le \alpha < \alpha_1$: coherent state. (ii) $\alpha_1 < \alpha < \alpha_2$: metastable transition between coherent state and partially synchronized state. (iii)$\alpha_2 < \alpha < \alpha_3$: partially synchronized state. (iv)$\alpha_3 < \alpha < \alpha_4$: metastable transition between partially synchronized state and modulated state. (v) $\alpha_4 \le \alpha < 0.5$: incoherent state.}
\label{fig:6}
\end{figure}

For $0.485\pi < \alpha < 0.495\pi$, the system shows a novel state we call the modulated state. This state is characterized by a spatially periodic modulation of the amplitude of the oscillators. The system again displays metastable behavior between the partially synchronized state and modulated state (Fig. ~\ref{fig:1}(e)).
We define the variance of the amplitude
\begin{eqnarray}
\chi &=& \left < V \right >_T, \\
V(t) &=& \frac{1}{M}\sum_{\sigma=1}^M \left (  r_\sigma(t)- \left < r_\sigma(t) \right >  \right )^2 ,
\end{eqnarray} 
where $\left < V \right >_T$ means the time average, to characterize these different states (it was also used in the previous study \cite{Shanahan2010Metastable} to evaluate the degree of chimera behavior in the system). Fig.~\ref{fig:6} shows the dependence of $\chi$ on $\alpha$. Time average is taken over $0 \ge t \ge 5000$, and 30 trials with different initial conditions for each value of $\alpha$ are shown. When $\alpha<\alpha_1\approx 0.44\pi$, the system rapidly settles into a stable coherent state with a constant $r$, and hence the variance of $r$ is almost zero. When $\alpha_1<\alpha<\alpha_2\approx 0.465\pi$, the system exhibits metastable behavior, transitioning back and forth between coherent and partially synchronized state (Fig.~\ref{fig:4}(b)), leading to an increase in $\chi$ as the proportion of time spent in the partially synchronized state grows. When $\alpha_2<\alpha<\alpha_3\approx 0.485\pi$, the system enters a persistent partially synchronized state, and the slow increase in $\chi$ is due to the increasing amplitude dispersion within the partially synchronized state. When $\alpha_3<\alpha<\alpha_4\approx 0.495\pi$, the system exhibits either a modulated state or metastable behavior between the modulated and partially synchronized states. In the latter case,  $\chi$ is larger than the partially synchronized state since the modulated state has a larger dispersion of $r$. This explains the branching behavior of $\chi$. Finally, when $\alpha_4 < \alpha \le 0.5\pi$, the amplitudes of all oscillators gradually decrease to zero.

\section{Conclusion}
We have studied the Ott-Antonsen reduced $M$-population Kuramoto-Sakaguchi oscillator model, and found a rich dynamical behavior by varying the phase lag parameter $\alpha$.
In contrast to the ordinary KS phase oscillator models \cite{maistrenko2014,Duguet2019,kawase2019critical,Li2021}, which show only the spatial coexistence of synchronized and unsynchronized states, we found the system exhibits novel metastable behavior.
In this system, the amplitude variable $r$ plays an important role, which makes the phase space of this system extremely complex, giving rise to the emergence of new attractors in the form of traveling waves.

Different from other amplitude-related behaviors such as amplitude chimeras\cite{Zakharova2014,Zakharova_2016},
this system continually transitions between different states, spending periods of time on the trajectory that is very close to the coherent state attractor (but not falling into its basin of attraction), and then at certain moments departing and becoming partially synchronizing for a duration, before again approaching the vicinity of the coherent state. 
These transitions between metastable states are not driven by external perturbations or noise, but rather by the inherent complex structure of the high-dimensional nonlinear system's phase space.
This metastability is attributed to the internal heterogeneity $\Delta$ within the population (setting $\Delta$ to 0 makes the mean-field behavior of the population equivalent to the original KS phase oscillator). 

Metastability, the coexistence of segregated and integrated neural dynamics, is recognized as a hallmark of brain activity in neuroscience~\cite{Emmanuelle2014the}. The brain's key functions are believed to be encoded by the coexistence of segregation (the tendency of neural ensembles to diverge and function independently) and integration (the tendency of neural ensembles to synchronize).
Furthermore, {a recent theoretical study suggested} 
a connection between the value of $\alpha$ and the ratio of electrical and chemical synapses 
in the brain~\cite{Clusella2022kuramoto}.
{According to the theory, the phase lag is close to $\pi/2$ if chemical synapses dominate
electrical ones in terms of the coupling strengths.
Although it is widely accepted that chemical synapses are much greater in number than the electrical ones~\cite{purves2007neuroscience},
precise estimates of their coupling strengths are lacking.
It will stimulate experimental studies to determine the inputs from the two types of 
synapses and estimate the phase lag. 
Also, it is desirable that other origins of the phase lag in neuronal networks are explored
and quantitatively measured.
Fine tuning of the phase lag will lead to verification of the metastable states found in the present work. 
}
By extending the OA-reduced $M$-population KS model to the brain network, we 
can link the microscopic and macroscopic descriptions and capture the microscopic properties of the underlying neuronal system. Given that 
the {KS}  model exhibits both the association with synaptic properties and metastable behavior, it is a promising candidate for brain network modeling.

\begin{acknowledgments}
This work was supported by JST, the establishment of university fellowships towards 
the creation of science technology innovation, Grant Number JPMJFS2102 to B. L.,
{JST SPRING, Grant Number JPMJSP2114 to B. L.,} 
and KAKENHI Grant No. JP21K03396 to N.U.

\end{acknowledgments}

\section*{Data Availability Statement}

The data that support the findings of this study are available from the corresponding author upon reasonable request.

\appendix

\section{Ott-Antonsen Reduction of $M$-population of KS model with heterogeneity}
\label{appendix_OA}
Here we show the derivation of the reduced dynamic equation (\ref{eq:m-population_KS_op_dynamics}) from the original KS model (\ref{eq:m-population_KS}).

Considering the thermodynamic limit where $N\to\infty$, the order parameter for each population can be defined as
\begin{equation}
    z_\sigma(t) = \int_{-\infty}^\infty \int_0^{2\pi} f_\sigma(\omega^\sigma, \theta^\sigma, t) e^{i\theta^\sigma} \mathrm{d} \theta^\sigma \mathrm{d} \omega^\sigma, \label{eq:m-population_KS_op_def}
\end{equation}
where $f_\sigma(\omega^\sigma, \theta^\sigma, t)$ is the probability density of oscillators in population $\sigma$, obeying the continuity equation
\begin{equation}
    \frac{\partial f_\sigma}{\partial t} + \frac{\partial}{\partial\theta^\sigma}(f_\sigma v_\sigma) = 0, \label{eq:m-population_KS_continuity_def}
\end{equation}
where $v_\sigma(\omega^\sigma, \theta^\sigma, t)$ is their velocity, 
{
\begin{eqnarray}
    v_\sigma &=& \omega^\sigma - \sum_{\tau=1}^M K_{\sigma\tau}\int_\infty^\infty \int_0^{2\pi} f_\tau(\omega^\tau,\theta^\tau,t) \nonumber\\
    &\times&\sin(\theta^\sigma-\theta^\tau+\alpha_{\sigma\tau})\mathrm{d} \theta^\tau \mathrm{d} \omega^\tau .
\end{eqnarray}
Noting that
\begin{equation} \sin(\theta^\sigma-\theta^\tau+\alpha_{\sigma\tau})=\frac{e^{i(\theta^\sigma-\theta^\tau+\alpha_{\sigma\tau})}-e^{-i(\theta^\sigma-\theta^\tau+\alpha_{\sigma\tau})}}{2i},
\end{equation}
the velocity can be written as
}
\begin{equation}
    v_\sigma = \omega^\sigma - \sum_{\tau=1}^M \frac{K_{\sigma\tau}}{2i}\left [ 
    e^{i(\theta^\sigma + \alpha_{\sigma\tau})} z_\tau^*(t) - e^{-i(\theta^\sigma + \alpha_{\sigma\tau})} z_\tau(t) \right ].  \label{eq:m-population_KS_continuity_v}
\end{equation}
Employing the Ott-Antonsen ansatz \cite{ott2008low}, $f_\sigma(\omega^\sigma, \theta^\sigma, t)$ can be expanded as a Fourier series in $\theta$ with coefficients $f_{\sigma,n}(\omega^\sigma,t)=[a(\omega^\sigma,t)]^n$,
\begin{equation}
    f_\sigma(\omega^\sigma, \theta^\sigma, t) = \frac{g_\sigma(\omega^\sigma)}{2\pi} \left \{ 1 + \sum_{n=1}^\infty [a_\sigma(\omega^\sigma, t)]^n e^{in\theta^\sigma} + c.c. \right \}. \label{eq:m-population_KS_fourier}
\end{equation}
Putting Eq.~(\ref{eq:m-population_KS_fourier}) into Eq.~(\ref{eq:m-population_KS_op_def}), we can write $z_\sigma(t)$ as a function of $a$,
\begin{equation}
    z_\sigma(t) = \int_{-\infty}^\infty g_\sigma(\omega^\sigma) [a_\sigma(\omega^\sigma, t)]^* \mathrm{d} \omega^\sigma.
    \label{eq:m-population_KS_op_OA}
\end{equation}
Putting Eq.~(\ref{eq:m-population_KS_fourier}) into Eq.~(\ref{eq:m-population_KS_continuity_def}), then extracting the coefficients of the 1st-order term $e^{i\theta^\sigma}$, we get
\begin{equation}
    \frac{\partial a_\sigma}{\partial t} = -i\omega^\sigma a_\sigma + \sum_{\tau=1}^M \frac{K_{\sigma\tau}}{2} \left [ e^{i\alpha_{\sigma\tau}} z_\tau^* - a_\sigma^2 e^{-i\alpha_{\sigma\tau}} z_\tau \right ].
    \label{eq:m-population_KS_a_dynamics}
\end{equation}
By choosing a Lorentzian distribution with center $\Omega_\sigma$ and width $\Delta_\sigma$,
\begin{equation}
    g_\sigma(\omega^\sigma) = \frac{1}{\pi}\frac{\Delta_\sigma}{(\omega^\sigma-\Omega_\sigma)^2+\Delta_\sigma^2},
    \label{eq:m-population_KS_Lorentzian_def}
\end{equation}
the integral in Eq.~(\ref{eq:m-population_KS_op_OA}) can be calculated by closing the contour by a large semicircle in the lower half $\omega^\sigma$-plane, where the integral is given by the residue of the pole at $\omega^\sigma=\Omega_\sigma-i\Delta_\sigma$,
\begin{equation}
    z_\sigma(t) = \left [a_\sigma(\Omega_\sigma-i\Delta_\sigma,t) \right]^*.
    \label{eq:m-population_KS_op_integrated}
\end{equation}
Finally, using Eq.~(\ref{eq:m-population_KS_a_dynamics}), the dynamics of the order parameter $z_\sigma(t)$ can be represented as
\begin{equation}
    \frac{\mathrm{d} z_\sigma^*}{\mathrm{d} t}=-(\Delta_\sigma + i\Omega_\sigma) z_\sigma^* + \sum_{\tau=1}^M \frac{K_{\sigma\tau}}{2} \left [ e^{i\alpha_{\sigma\tau}} z_\tau^* - e^{-i\alpha_{\sigma\tau}} z_\tau (z_\sigma^*)^2 \right ].
    \label{eq:m-population_KS_op_dynamics}
\end{equation}  
This equation can be rewritten in the polar coordinates defined via $z_\sigma=r_\sigma e^{-i\phi_\sigma}$ where $r_\sigma$ and $\phi_\sigma$ are the radial and angular components of the order parameters, respectively. 
\begin{align}
    \frac{\mathrm{d} r_\sigma}{\mathrm{d} t} &=
    -\Delta_\sigma r_\sigma 
    + \frac{1-r_\sigma^2}{2} \sum_{\tau=1}^M K_{\sigma\tau} r_\tau \cos(\alpha_{\sigma\tau}+\phi_\tau-\phi_\sigma),
\label{drdt}
\\
\frac{\mathrm{d} \phi_\sigma}{\mathrm{d} t} &= 
-\Omega_\sigma + \frac{1+r_\sigma^2}{2r_\sigma} \sum_{\tau=1}^M K_{\sigma\tau} r_\tau \sin(\alpha_{\sigma\tau}+\phi_\tau-\phi_\sigma). 
\label{dphidt}
\end{align}

\section{Linear Stability Analysis}
\label{appendix_LSA}
{
Let us denote the fixed point solutions in Eqs.(\ref{eq:5})-(\ref{eq:7}) as 
$r_{\textsf{fp}}$ and $\phi_{\textsf{fp}}$.}
Substituting $r_\sigma = {r_{\textsf{fp}}} +\delta r_\sigma$ and 
$\phi_\sigma = {\phi_{\textsf{fp}}} + \delta \phi_\sigma$ 
into Eqs.~(\ref{eq:r})(\ref{eq:phi})  
and linearizing them with respect to the {small perturbations} $\delta r_\sigma, \delta \phi_\sigma$, we obtain
\begin{eqnarray}
f_\sigma &=& \frac{\mathrm{d} }{\mathrm{d} t} \delta r_\sigma \nonumber \\
&=& \left ( -\Delta-Kr_\sigma\sum_{\tau=\sigma-R}^{\sigma+R} r_\tau \cos(\alpha+\phi_\tau-\phi_\sigma)  \right )\delta r_\sigma  \nonumber \\
&+&  \frac{1-r_\sigma^2}{2}K\sum_{\tau=\sigma-R}^{\sigma+R} \cos(\alpha+\phi_\tau-\phi_\sigma)   \delta r_{\tau}\nonumber\\
&+& \frac{1-r_\sigma^2}{2}K\sum_{\tau=\sigma-R}^{\sigma+R} r_\tau \sin(\alpha+\phi_\tau-\phi_\sigma) \delta\phi_\sigma\nonumber\\
&-&   \frac{1-r_\sigma^2}{2}K\sum_{\tau=\sigma-R}^{\sigma+R} r_\tau \sin(\alpha+\phi_\tau-\phi_\sigma)  \delta\phi_{\tau},
\label{fsigma}
\\
g_\sigma &=& \frac{\mathrm{d} }{\mathrm{d} t} \delta \phi_\sigma \nonumber \\
&=& \frac{r_\sigma^2-1}{2r_\sigma^2} K\sum_{\tau=\sigma-R}^{\sigma+R} r_\tau \sin(\alpha+\phi_\tau-\phi_\sigma)  \delta r_\sigma  \nonumber \\
&+& \frac{1+r_\sigma^2}{2r_\sigma}K\sum_{\tau=\sigma-R}^{\sigma+R} \sin(\alpha+\phi_\tau-\phi_\sigma) \delta r_{\tau}\nonumber\\
&-& \frac{1+r_\sigma^2}{2r_\sigma}K\sum_{\tau=\sigma-R}^{\sigma+R} r_\tau \cos(\alpha+\phi_\tau-\phi_\sigma) \delta\phi_\sigma\nonumber\\
&+&  \frac{1+r_\sigma^2}{2r_\sigma}K\sum_{\tau=\sigma-R}^{\sigma+R} r_\tau \cos(\alpha+\phi_\tau-\phi_\sigma)  \delta\phi_{\tau}. 
\label{gsigma}
\end{eqnarray}
{Note that the $O(\delta \phi)$ terms are obtained by 
expanding the cosine and sine terms in Eqs.(\ref{drdt}),(\ref{dphidt}).
}
{
We rewrite Eqs.(\ref{fsigma}), (\ref{gsigma}) in the matrix form
 $(f_1, g_1, f_2, g_2, \ldots, f_M, g_M)^{\rm T} = {\cal C} \cdot 
 (
\delta r_1, \delta \phi_1,
\delta r_2, \delta \phi_2, \ldots, 
\delta r_M, \delta \phi_M)^{\rm T}$, where the
coefficient matrix ${\cal C}$ is a $2M \times 2M$ matrix.
}
According to the Bloch theorem, the $2M$ eigenvalues are identical to those of the following $M$ small $2\times 2$ matrices:
\begin{eqnarray}
&&
\begin{pmatrix}
 \frac{\partial f_\sigma}{\partial r_\sigma} +  \frac{\partial f_\sigma}{\partial r_\tau}e^{i\frac{2\pi}{M}m } & \frac{\partial f_\sigma}{\partial \phi_\sigma}+\frac{\partial f_\sigma}{\partial \phi_\tau}e^{i\frac{2\pi}{M}m }\\
\frac{\partial g_\sigma}{\partial r_\sigma}+\frac{\partial g_\sigma}{\partial r_\tau}e^{i\frac{2\pi}{M}m }  & \frac{\partial g_\sigma}{\partial \phi_\sigma}+\frac{\partial g_\sigma}{\partial \phi_\tau}e^{i\frac{2\pi}{M}m }
\end{pmatrix} \nonumber \\
&=&
\begin{pmatrix}
 \Delta-Kh\cos\alpha + \Delta e^{i\frac{2\pi}{M}m } & \Delta r \tan \alpha-\Delta r \tan\alpha e^{i\frac{2\pi}{M}m }\\
 -\frac{\Delta\tan\alpha}{r} +\frac{\sin\alpha-\Delta\tan\alpha}{r}e^{i\frac{2\pi}{M}m } &\Delta-1+(1-\Delta)e^{i\frac{2\pi}{M}m } 
\end{pmatrix} \nonumber\\
&=&
\begin{pmatrix}
 a_m & b_m\\
 c_m & d_m 
\end{pmatrix}, 
\end{eqnarray}
where $m = 1,2,\dots,M$. Therefore, the eigenvalues are:
\begin{eqnarray}
\lambda _{m\pm} = \frac{(a_m+d_m)\pm\sqrt{(a_m+d_m)^2-4(a_m d_m-b_m c_m)} }{2}. 
\end{eqnarray}
Then we extract the real part of the eigenvalues as
\begin{eqnarray}
\mathrm{Re}(\lambda_{m\pm}) = \frac{ C_0 + \cos\left(\frac{2\pi}{M}m\right) \pm 
(x^2+y^2)^\frac{1}{4}\cos \left ( \frac{1}{2}\tan^{-1} (x,y)  \right )
} {2}, 
\nonumber\\
\end{eqnarray}
where
\begin{eqnarray}
x &=& C_1 + C_2\cos\left(\frac{2\pi}{M}m\right) + C_3 \cos\left(\frac{4\pi}{M}m\right), 
\\
y &=& C_2\sin\left(\frac{2\pi}{M}m\right) + C_3 \sin\left(\frac{4\pi}{M}m\right),
\\
C_0 &=& 2\Delta - 1 - Kh\cos\alpha,  
\\
C_1 &=& (A\cos\alpha-1)^2-4\Delta^2\tan^2\alpha, 
\\
C_2 &=& (A\cos\alpha-1)(2-4\Delta)+4\Delta\sin\alpha\tan\alpha,  
\\
C_3 &=& (2\Delta-1)^2+4(\Delta^2 \tan ^2\alpha-\Delta\sin\alpha\tan\alpha). 
\end{eqnarray}

\nocite{*}
%\bibliography{li2024metastability}% Produces the bibliography via BibTeX.
\providecommand{\noopsort}[1]{}\providecommand{\singleletter}[1]{#1}%

\end{document}